\documentclass[submission,copyright]{eptcs}
\providecommand{\event}{TARK 2023} % Name of the event you are submitting to

\usepackage{iftex}

\ifpdf
  \usepackage{underscore}         % Only needed if you use pdflatex.
  \usepackage[T1]{fontenc}        % Recommended with pdflatex
\else
  \usepackage{breakurl}           % Not needed if you use pdflatex only.
\fi

\usepackage{amsmath}
\usepackage{amsfonts}
\usepackage{amssymb}
\usepackage{stmaryrd}
\usepackage{array}
\usepackage[pdftex]{graphicx}
\usepackage{soul}
\usepackage[english]{babel}
\usepackage{multicol}

\newtheorem{ass}{Assumption}

\newtheorem{cor}{Corollary}
\newtheorem{defin}{Definition}

\newtheorem{lem}{Lemma}

\newtheorem{prop}{Proposition}

\newcommand{\w}{\omega}

\begin{document}
	
\title{Implicit Knowledge in Unawareness Structures \\ - Extended Abstract -\thanks{Dedicated to Joe Halpern on the occasion of his 70th birthday. A complete version of the paper with proofs of the results is available from \url{https://faculty.econ.ucdavis.edu/faculty/schipper/implicit.pdf}. We thank three anonymous reviewers, Hans van Ditmarsch, participants in the CSLI Workshop 2022, Stanford University, and the 2022 Zoom Mini-Workshop on Epistemic Logic with Unawareness for helpful comments. Gaia gratefully acknowledges funding from the Carlsberg Foundation. Burkhard gratefully acknowledges financial support via ARO Contract W911NF2210282.}}

\author{Gaia Belardinelli
\institute{Center for Information and Bubble Studies\\ University of Copenhagen}
\email{belardinelli@hum.ku.dk}
\and Burkhard C. Schipper
\institute{Department of Economics\\ University of California, Davis}
\email{bcschipper@ucdavis.edu}}
\def\titlerunning{Implicit Knowledge in Unawareness Structures}
\def\authorrunning{Gaia Belardinelli \& Burkhard C. Schipper}

\maketitle
	
\begin{abstract} Awareness structures by Fagin and Halpern (1988) (FH) feature a syntactic awareness correspondence and accessibility relations modeling implicit knowledge. They are a flexible model of unawareness, and best interpreted from a outside modeler's perspective. Unawareness structures by Heifetz, Meier, and Schipper (2006, 2008) (HMS) model awareness by a lattice of state-spaces and explicit knowledge via a possibility correspondence. They can be interpreted as providing the subjective views of agents. Open questions include (1) how implicit knowledge can be defined in HMS structures, and (2) in which way FH structures can be extended to model the agents' subjective views. In this paper, we address (1) by showing how to derive implicit knowledge from explicit knowledge in HMS models. We also introduce a variant of HMS models that instead of explicit knowledge, takes implicit knowledge and awareness as primitives. Further, we address (2) by introducing a category of FH models that are modally equivalent relative to sublanguages and can be interpreted as agents' subjective views depending on their awareness. These constructions allow us to show an equivalence between HMS and FH models. As a corollary, we obtain soundness and completeness of HMS models with respect to the Logic of Propositional Awareness, based on a language featuring \emph{both} implicit and explicit knowledge.
\newline
\newline
\textbf{JEL-Classifications:} D83, C70.
\bigskip
\end{abstract}

\section{Introduction}
	
Models of unawareness are of interest in various disciplines, most notably in computer science, economics, game theory, decision theory, 
and philosophy. The seminal contribution in computer science and philosophy are awareness structures by Fagin and Halpern (1988) (henceforth, \emph{FH models}) who extended Kripke structures with a syntactic awareness correspondence in order to feature notions of implicit knowledge, explicit knowledge, and awareness. In economics, Heifetz, Meier, and Schipper (2006, 2008) introduced unawareness structures (henceforth, \emph{HMS models}) that consist of a lattice of state spaces featuring a notion of explicit knowledge and awareness. Like Kripke structures, HMS models can be constructed canonically and three different sound and complete axiomatizations have been presented (Halpern and R\^{e}go, 2008, Heifetz, Meier, and Schipper, 2008).\footnote{\scriptsize For other approaches and an overview, see Schipper (2015).} There have already been a number of applications to game theory, decision theory, mechanism design and contracting, financial markets, electoral campaigning, conflict resolution, social network formation, business strategy and entrepreneurship etc.
	
The different modeling approaches may be seen as reflecting the different foci of the fields. HMS models in economics are very much motivated by game theory and its applications. The main underlying idea is that \emph{explicit} rational reasoning of players is what drives their behavior. Hence, the model features only explicit knowledge (without the detour via implicit knowledge) and awareness, and it can be interpreted as encompassing the subjective views of players. Moreover, the syntax-free frame lends itself seamlessly to the existing body of work in decision theory and game theory. The focus on behavioral implications also explains why the model is built on strong properties of knowledge such as (positive) introspection and factivity: this allows for the identification of the behavioral implications of unawareness per se without confounding it with mistakes in information processing. Somewhat differently, FH models were motivated more generally by the study of the logical non-omniscience problem in computer science and philosophy (see e.g., Hintikka, 1975, Levesque, 1984, Lakemeyer, 1986, Stalnaker, 1991). They represent awareness via syntactic awareness correspondences, which for each agent assigns a set of formulas to each state. This approach to awareness modeling offers a great deal of flexibility, because the set of formulas an agent is aware of may be arbitrary, thereby allowing potentially for the representation of different notions of awareness.\footnote{\scriptsize See Fagin and Halpern (1988, pp. 54-55) and Fagin et al. (1995, Chapter 9.5) for discussions.} However, because their semantics is not syntax-free, their applications to decision or game theory require more effort. This is because in decision theory and game theory and applications thereof, the primitives are typically not described syntactically. Moreover, FH models are best interpreted as a tool used by an outside modeler (like a systems designer of a multi-agent distributed system) for two reasons: First, the primitive notion of knowledge is \emph{implicit} knowledge while explicit knowledge is derived from implicit knowledge and awareness. Implicit knowledge is not necessarily something that the agent herself can consciously reason about. Second, we cannot think of FH models as models that the agents themselves use for analyzing their epistemic universe unless they are aware of everything.\footnote{\scriptsize The view of the systems designer is expressed eloquently by Fagin, Halpern, and Vardi (1986): ``The notion of knowledge is \emph{external}. A process cannot answer questions based on its knowledge with respect to this notion of knowledge. Rather, this is a notion meant to be used by the system designer reasoning about the system. ... (I)t does seem to capture the type of intuitive reasoning that goes on by system designers.''} This becomes relevant in interactive settings when we are interested in the players' interactive perceptions of the epistemic universe. Despite the differences in modeling approaches, Halpern and R\^{e}go (2008) and Belardinelli and Rendsvig (2022) formalize in which ways HMS models are equivalent to FH models in terms of explicit knowledge and awareness. However, as the discussion above makes clear, there remain open questions: First, can implicit knowledge be captured also in HMS models and how would this notion of implicit knowledge be related to implicit knowledge in FH models? Second, can we extend FH models so as to interpret them from the agents' subjective point of views? These questions will be answered in this paper. 

By showing how to derive implicit knowledge from explicit knowledge in HMS models, we provide a way to understand implicit knowledge in terms of explicit knowledge. We are aware of only a few other approaches deriving implicit knowledge from explicit knowledge. Using neighborhood models without a notion of awareness, Vel\'{a}zquez-Quesada (2013) takes explicit knowledge as the primitive and then derives implicit knowledge as closure of logical consequences of explicit knowledge. Implicit knowledge is then understood as knowledge that the agent ideally could deduce from her explicit knowledge. Lorini (2020) takes an agent's belief base as explicit knowledge and derives implicit knowledge as what is deducible from an agent's belief base and common background information. While we find these two notions of implicit knowledge easy to interpret, it is not the notion of implicit knowledge that is captured by propositionally determined FH models, namely models where an agent is aware of a formula if and only if she is aware of all atomic formula that appear in it. We also introduce a variant of HMS models in which we take the notion of implicit knowledge and a semantic awareness function as the primitive and then derive explicit knowledge. This shows that in HMS models, implicit and explicit knowledge are ``interdefinable'', at least in the sense that taking any of the two as primitive is sufficient to recover the other, so one may choose either one as primitive. 

We are also interested in an extension of FH models that allows us to interpret them as subjective views of agents. Starting from an FH model, we show how to form a category of FH models with FH models as objects and surjective bounded morphisms as morphisms. Each category of FH models is literally a category of FH models that are modally equivalent relative to sublanguages formed by taking subsets of atomic formulae. The category of FH models forms a complete lattice ordered by subset inclusion on sets of atomic formulae or ordered by surjective bounded morphisms or ordered by modal equivalence relative to sublanguages. Each FH model in the lattice can be interpreted as the subjective model of an agent with that awareness level given by the subset of atomic formulae for which the FH model is defined. The construction now suggests transformations between FH and HMS models. The transformation from FH to HMS models relies on a transformation of each FH category into an implicit knowledge-based HMS model mentioned above. This implicit knowledge-based HMS model can be complemented with explicit knowledge and thus yields a HMS model. The transformation from HMS to FH models simply relies on pruning away the subjective spaces, only maintaining the upmost space in the lattice, as well as deriving the syntactic awareness correspondences from possibility correspondences and the lattice of spaces in HMS models. For each model class, its transformation into a model of the other class satisfy the same formulas from a language for explicit, implicit knowledge, and awareness. It shows how the model classes and implicit knowledge notions relate to each other. As a corollary of soundness and completeness of the Logic of Propositional Awareness w.r.t. the class of FH models, the results allow us to derive soundness and completeness for the class of HMS models with implicit knowledge, complementing earlier axiomatizations of HMS models that made use of explicit knowledge (and awareness) only (Halpern and R\^{e}go, 2008, Heifetz, Meier, and Schipper, 2008).

\section{HMS Models}
	
HMS models are multi-agent models for awareness originally proposed by Heifetz, Meier, and Schipper (2006, 2008). For lack of space, we refer the reader for explanations and intuitions to that papers. Throughout the paper, we let $\mathsf{At}$ be a non-empty set of atomic formulas.
	
\begin{defin}\label{HMS_model}  A \emph{HMS model} $\mathsf{M} = \langle I, \{S_{\Phi}\}_{\Phi \subseteq \mathsf{At}}, (r^{\Phi}_{\Psi})_{\Psi \subseteq \Phi \subseteq \mathsf{At}}, (\Pi_i)_{i \in I}, v \rangle$ for $\mathsf{At}$ consists of

\begin{itemize}
			
	\item a non-empty set of individuals $I$,
			
	\item a non-empty collection of non-empty disjoint state spaces $\{S_{\Phi}\}_{\Phi \subseteq \mathsf{At}}$ indexed by subsets of atomic formulae $\Phi \subseteq \mathsf{At}$. Note that $\{S_{\Phi}\}_{\Phi \subseteq \mathsf{At}}$ forms a complete lattice by subset inclusion on the set of atomic formulae $\Phi \subseteq \mathsf{At}$. Denote the set of all states in spaces of the lattice by $\Omega := \bigcup_{\Phi \subseteq \mathsf{\mathsf{At}}} S_{\Phi}$.
			
	\item Projections $(r^\Phi_\Psi)_{\Psi \subseteq \Phi \subseteq \mathsf{At}}$ such that, for any $\Phi, \Psi \subseteq \mathsf{\mathsf{At}}$ with $\Psi \subseteq \Phi$, $r^\Phi_\Psi: S_{\Phi} \longrightarrow S_{\Psi}$ is surjective, for any $\Phi \subseteq \mathsf{\mathsf{At}}$, $r^{\Phi}_{\Phi}$ is the identity on $S_{\Phi}$, and for any $\Phi, \Psi, \Upsilon \subseteq \mathsf{\mathsf{At}}$, $\Upsilon \subseteq \Psi \subseteq \Phi$, $r^{\Phi}_{\Upsilon} = r^{\Psi}_{\Upsilon} \circ r^{\Phi}_{\Psi}$. For any $\Phi \subseteq \mathsf{At}$ and $D \subseteq S_{\Phi}$, denote by $D^{\uparrow} := \bigcup_{\Phi \subseteq \Psi \subseteq \mathsf{At}} (r^{\Psi}_{\Phi})^{-1}(D)$. An \emph{event} $E \subseteq \Omega$ is defined by a $\Phi \subseteq \mathsf{At}$ and a subset $D \subseteq S_{\Phi}$ such that $E := D^{\uparrow}$. We call $S_{\Phi}$ the base-space of the event $E$ and $D$ the base of the event $E$. We denote by $\Sigma$ the set of events.
			
	\item A possibility correspondence $\Pi_i: \Omega \longrightarrow 2^{\Omega} \setminus \{\emptyset\}$ for each individual $i \in I$.
			
	\item A valuation function $v: \mathsf{At} \longrightarrow \Sigma$.
			
\end{itemize}
\end{defin}
	
Not every subset of the union of spaces is an event. Intuitively, $D^\uparrow$ collects all the ``extensions of descriptions in $D$ to at least as expressive vocabularies" (Heifetz, Meier, and Schipper, 2006). Events are well defined by the above definition except for the case of vacuous events. Since the empty set is a subset of any space, we have as many vacuous events as there are state-spaces. These vacuous events are distinguished by their base-space, so we denote them by $\emptyset^{S_{\Phi}}$ for $\Phi \subseteq \mathsf{At}$. At a first glance, the existence of many vacuous events may be puzzling. Note that vacuous events essentially represent contradictions, i.e., propositions that cannot hold at any state. Contradictions are formed with atomic formulae. Thus, they can be more or less complicated depending on the expressiveness of the underlying language describing states and hence are represented by different vacuous events.
	
We define Boolean operations on events. Negation of events is defined as follows: Let $E$ be an event with base $D$ and base-space $S_{\Phi}$. Then $\neg E := (S_{\Phi} \setminus D)^{\uparrow}$. Conjunction of events is defined by intersection of events. Disjunction of events is defined by the DeMorgan Law using negation and conjunction as just defined. Note that in HMS models we typically have $E \cup \neg E \subsetneqq \Omega$ unless the base-space of $E$ is $S_{\emptyset}$, the meet of the lattice of spaces. Also, disjunction of two events is typically a proper subset of the union of these events unless both events have the same base-space, since it is just the union of the events in spaces in which both events are ``expressible''.
	
The following notation will be convenient: Sometimes we denote by $S_{\omega}$ the state-space that contains state $\omega$.  For any $D \subseteq S_{\Phi}$, we denote by $D_{S_{\Psi}}$ the projection of $D$ to $S_{\Psi}$ for $\Psi \subseteq \Phi \subseteq \mathsf{At}$. We simplify notation further and let for any $D \subseteq S_{\Phi}$ and $\Psi \subseteq \Phi \subseteq \mathsf{At}$, $D_{\Psi}$ be the projection of $D$ to $S_{\Psi}$. Similarly, for any $D \subseteq S_{\Psi}$, we denote by $D^{\Phi}$ the ``elaboration'' of $D$ in the space $S_{\Phi}$ with $\Psi \subseteq \Phi$, i.e., $D^{\Phi} := (r^{\Phi}_{\Psi})^{-1}(D)$. The same applies to states, i.e., $\omega_{\Psi}$ is the projection of $\omega \in S_{\Phi}$ to $S_{\Psi}$ with $\Psi \subseteq \Phi$. Finally, for any event $E \in \Sigma$, we denote by $S(E)$ the base-space of $E$. We say that an event $E$ is expressible in $S_\Phi$ if $S(E) \preceq S_\Phi$. 
	
As usual in epistemic structures used in game theory and economics, information is modeled by a possibility correspondence instead of an accessibility relation. In HMS models, having mappings rather than relations adds extra convenience in that we can easily compose projections with possibility correspondences and vice versa. It is precisely the projective structure that makes HMS models tractable in applications and lets us analyze phenomena across ``awareness levels'' $\{S_{\Phi}\}_{\Phi \subseteq \mathsf{At}}$. Since the motivation for HMS models in game theory and economics is to isolate the behavioral implications of unawareness from other factors like mistakes in information processing etc., we require the possibility correspondences satisfy strong properties analogous to S5.\footnote{\scriptsize Again, for lack of space we refer to Heifetz, Meier, and Schipper (2006, 2008) for discussions of these properties. Generalizations are considered by Heifetz, Meier, and Schipper (2013a), Halpern and R\^{e}go (2008), Board, Chung, and Schipper (2011), and Galanis (2011, 2013).} 
	
\begin{ass}\label{assumptions_HMS} For any individual $i \in I$, we require that the possibility correspondence $\Pi_i$ satisfies
\begin{itemize}

\item[] \emph{Confinement:} If $\omega \in S_{\Phi}$, then $\Pi_{i}(\omega)\subseteq
			S_{\Psi}$ for some $\Psi \subseteq \Phi$.
			
\item[] \emph{Generalized Reflexivity:} $\omega \in \Pi_{i}^{\uparrow}(
			\omega)$ for every $\omega \in \Omega$.\footnote{\scriptsize Here and in what follows, we abuse notation slightly and write $\Pi_{i}^{\uparrow }(\omega)$ for $\left(\Pi _{i}(\omega )\right)^{\uparrow }$.}
			
\item[] \emph{Stationarity:} $\omega' \in \Pi_{i}(\omega)$
			implies $\Pi_{i}(\omega') =\Pi_{i}(\omega)$.
			
\item[] \emph{Projections Preserve Ignorance:} If $\omega \in S_{\Phi}$ and $\Psi \subseteq \Phi$, then $\Pi_{i}^{\uparrow}(\omega) \subseteq \Pi_{i}^{\uparrow}(\omega_{\Psi})$.
			
\item[] \emph{Projections Preserve Knowledge:} If $\Upsilon \subseteq \Psi \subseteq \Phi$, $\omega \in S_{\Phi}$ and $\Pi_{i}(\omega)\subseteq S_{\Psi}$ then $\left(\Pi_{i}( \omega)\right)_{\Upsilon} = \Pi_{i}(\omega_{\Upsilon})$.
\end{itemize}
\end{ass}

Sometimes we denote by $S_{\Pi_i(\omega)}$ the state-space $S$ for which $\Pi_i(\omega) \subseteq S$. We refer to Heifetz, Meier, and Schipper (2006, 2008) for discussions of these properties.

Given the possibility correspondence, the knowledge operator is defined as usual
\begin{defin}\label{definK} For every individual $i \in I$, the \emph{knowledge operator} on events is defined by, for every event $E \in \Sigma$, $K_i(E) := \{\omega \in \Omega : \Pi_i(\omega) \subseteq E \}$ if there exists a state $\omega \in \Omega$ such that   $\Pi_i(\omega) \subseteq E$, and by $K_i(E) = \emptyset^{S(E)}$ otherwise.
\end{defin} 
\begin{defin}\label{definA} For every individual $i \in I$, the \emph{awareness operator} on events is defined by, for every event $E \in \Sigma$, $A_i(E) := \{\omega \in \Omega : S_{\Pi_i(\omega)} \succeq S(E) \}$ if there exists a state $\omega \in \Omega$ such that $S_{\Pi_i(\omega)} \succeq S(E)$, and by $A_i(E) = \emptyset^{S(E)}$ otherwise. The unawareness operator is defined by $U_i(E) := \neg A_i(E)$.
\end{defin}

We read $K_i(E)$ as ``individual $i$ knows the event $E$'' and $A_i(E)$ as ``individual $i$ is aware of event $E$''. 
	
\begin{lem}[Heifetz, Meier, and Schipper, 2006]\label{alles_events} For every individual $i \in I$ and event $E \in \Sigma$, both $K_i(E)$ and $A_i(E)$ are $S(E)$-based events.
\end{lem}
	
\begin{prop}[Heifetz, Meier, and Schipper, 2006]\label{strong_explicit_knowledge} For every individual $i \in I$, the knowledge operator $K_i$ satisfies the following properties: For every $E, F \in \Sigma$ and $\{E_n\}_n \subseteq \Sigma$,
\begin{itemize}
\item[(i)] Necessitation: $K_{i}(\Omega) = \Omega$,
			
\item[(ii)] Conjunction: $K_{i}\left( \bigcap_{n} E_{n}\right) = \bigcap_{n}K_{i}\left(E_{n}\right)$,
			
\item[(iii)] Truth: $K_{i}(E)\subseteq E$,
			
\item[(iv)] Positive Introspection: $K_{i}(E)\subseteq K_{i}K_{i}(E)$,
		
\item[(v)] Monotonicity: $E \subseteq F$ implies $K_{i}(E)\subseteq K_{i}(F)$.
			
\item[(vi)] Weak Negative Introspection I: $\neg K_{i}(E)\cap \neg K_{i} \neg K_{i}(E) \subseteq \neg K_{i} \neg K_{i} \neg K_{i}(E)$.
\end{itemize}
\end{prop}
	
\begin{prop}[Heifetz, Meier, and Schipper, 2006] For every individual $i \in I$, the following properties of knowledge and awareness obtain: For every $E \in \Sigma$ and $\{E_n\}_n \subseteq \Sigma$,
\begin{enumerate}
\item $KU$ Introspection: $K_{i}U_{i}(E) = \emptyset ^{S(E)}$,
			
\item $AU$ Introspection: $U_{i}(E) = U_{i}U_{i}(E)$
			
\item Weak Necessitation: $A_{i}(E) = K_{i}(S(E)^{\uparrow})$,
			
\item Plausibility: $A_{i}(E) = K_{i}(E) \cup K_{i} \lnot K_{i}(E)$,
			
\item Strong Plausibility: $U_{i}(E) = \bigcap_{n=1}^{\infty}\left(\lnot K_{i}\right)^{n}(E)$,
			
\item Weak Negative Introspection II: $\lnot K_{i}(E)\cap A_{i}\lnot K_{i}(E) = K_{i}\lnot K_{i}(E)$,
			
\item Symmetry: $A_{i}(E) = A_{i}(\lnot E)$,
			
\item $A$-Conjunction: $\bigcap_{n} A_{i}\left(E_{n}\right) = A_{i}\left(\bigcap_{n} E_{n}\right)$,
			
\item $AK$-Self Reflection: $A_{i}(E) = A_{i}K_{i}(E)$,
			
\item $AA$-Self Reflection: $A_{i}(E) = A_{i}A_{i}(E)$,
			
\item $A$-Introspection: $A_{i}(E) = K_{i}A_{i}(E)$.
\end{enumerate}
\end{prop}
	
The following lemma turns out to be very useful but has not been proved in the literature. (For the proof, see the full version of the paper.)
	
\begin{lem}\label{lemma: pc in comparable spaces are equal} For every individual $i \in I$ and any $\Upsilon \subseteq \Psi \subseteq \Phi \subseteq \mathsf{At}$, if $\omega \in S_{\Phi}$ and $\Pi_i(\omega) \subseteq S_\Upsilon$, then $\Pi_i(\omega_\Psi) = \Pi_i(\omega)$.
\end{lem}

\section{From Explicit to Implicit\label{explicit-to-implicit}}
	
In this section, we introduce the implicit possibility correspondence $\Lambda_i$ as derived from $\Pi_i$. We then define implicit knowledge as based on $\Lambda_i$ and show that it satisfies standard S5 properties as well as properties of Fagin and Halpern (1988) that are jointly satisfied by implicit knowledge, explicit knowledge, and awareness. 
	
From now on we call for any individual $i \in I$, $\Pi_i$ the \emph{explicit} possibility correspondence, $\Pi_i(\w)$ \emph{explicit} possibility set at $\omega$, and $K_i(E)$ the event that $i$ \emph{explicitly} knows $E$.
	
\begin{defin} Given the explicit possibility correspondence $\Pi_i$ of individual $i \in I$, let the \emph{implicit possibility correspondence} $\Lambda_i: \Omega \longrightarrow 2^\Omega$ satisfy
\begin{itemize}
\item[] \emph{Reflexivity:} For any $\omega \in \Omega$, $\omega \in \Lambda_i(\omega)$.

\item[] \emph{Stationarity}: $\omega' \in \Lambda_i(\omega)$ implies $\Lambda_i(\omega') = \Lambda_i(\omega)$.

\item[] \emph{Projections Preserve Implicit Knowledge:} For any $\Phi \subseteq \mathsf{At}$, if $\omega \in S_{\Phi}$, then $\Lambda_i(\omega)_{\Psi} = \Lambda_i(\omega_{\Psi})$ for all $\Psi \subseteq \Phi$.

\item[] \emph{Explicit Measurability:} $\omega' \in \Lambda_i(\omega)$ implies $\Pi_i(\omega') = \Pi_i(\omega)$.

\item[] \emph{Implicit Measurability:} $\omega' \in \Pi_i(\omega)$ implies $\Lambda_i(\omega') = \Lambda_i(\omega)_{S_{\Pi_i(\omega)}}$.
\end{itemize} Given an HMS model $M = \langle I, \{S_{\Phi}\}_{\Phi \subseteq \mathsf{At}},(r^{\Phi}_{\Psi})_{\Psi \subseteq \Phi \subseteq \mathsf{At}}, (\Pi_i)_{i \in I}, v \rangle$ and a collection of implicit possibility correspondences $(\Lambda_i)_{i \in I}$ satisfying the above properties, we call $\overline{\mathsf{M}} = \langle I, \{S_{\Phi}\}_{\Phi \subseteq \mathsf{At}},(r^{\Phi}_{\Psi})_{\Psi \subseteq \Phi \subseteq \mathsf{At}}, (\Pi_i)_{i \in I}, \\ (\Lambda_i)_{i \in I}, v \rangle$ a \emph{complemented HMS model}.
\end{defin}

A complemented HMS model is a HMS model complemented with implicit possibility correspondences for each individual. In the following, we discuss and derive some properties of the implicit possibility correspondence. It also demonstrates ways in which the implicit possibility correspondence is consistent with the explicit possibility correspondence. 

Reflexivity and Stationarity are standard and imply that $\{\Lambda_i(\omega)\}_{\omega \in S_{\Phi}}$ forms a partition of $S_{\Phi}$ for every $\Phi \subseteq \mathsf{At}$. It is straightforward to see that they also imply a strengthening of Confinement (Assumption~\ref{assumptions_HMS}): The implicit possibility set at a state must be a subset of the state's space. That is, both the state and the implicit possibility set are described using the \emph{same} language. More formally:

\begin{lem}[Strong Confinement] For any individual $i \in I$, $\Phi \subseteq \mathsf{At}$, and $\omega \in S_{\Phi}$, $\Lambda_i(\omega) \subseteq S_{\Phi}$.
\end{lem}

Projections Preserve Implicit Knowledge is analogous to Projections Preserve Knowledge satisfied by $\Pi_i$. The absence of Projections Preserve (Implicit) Ignorance from the above list of imposed properties may look puzzling at the first glance. Yet, as we show below it is implied by Strong Confinement and Projections Preserve Implicit Knowledge.
	
\begin{lem}[Projections Preserve Implicit Ignorance] For any individual $i \in I$, if $\Lambda_i$ satisfies Strong Confinement and Projections Preserve Implicit Knowledge, then $\Lambda_i$ satisfies Projections Preserve Implicit Ignorance. That is, for all $\Phi \subseteq \mathsf{At}$, if $\omega \in S_{\Phi}$, then $\Lambda_i^{\uparrow}(\omega) \subseteq \Lambda_i^{\uparrow}(\omega_{\Psi})$ for all $\Psi \subseteq \Phi$.
\end{lem}
	
\begin{figure}\caption{Examples of Implicit Knowledge in Unawareness Structures\label{examples}}
\begin{center}
\includegraphics[scale = 0.08]{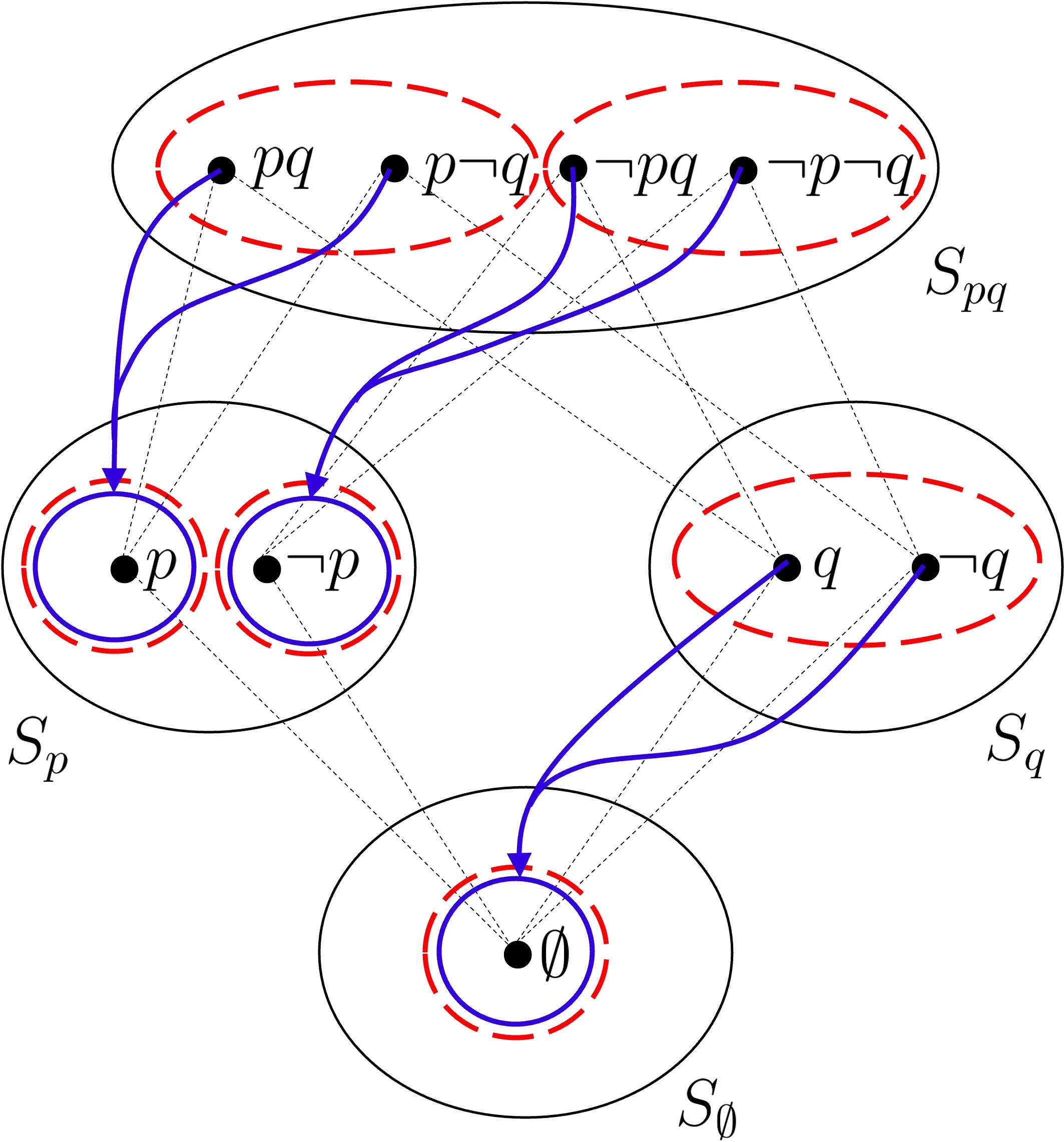} \quad \quad \quad \includegraphics[scale = 0.08]{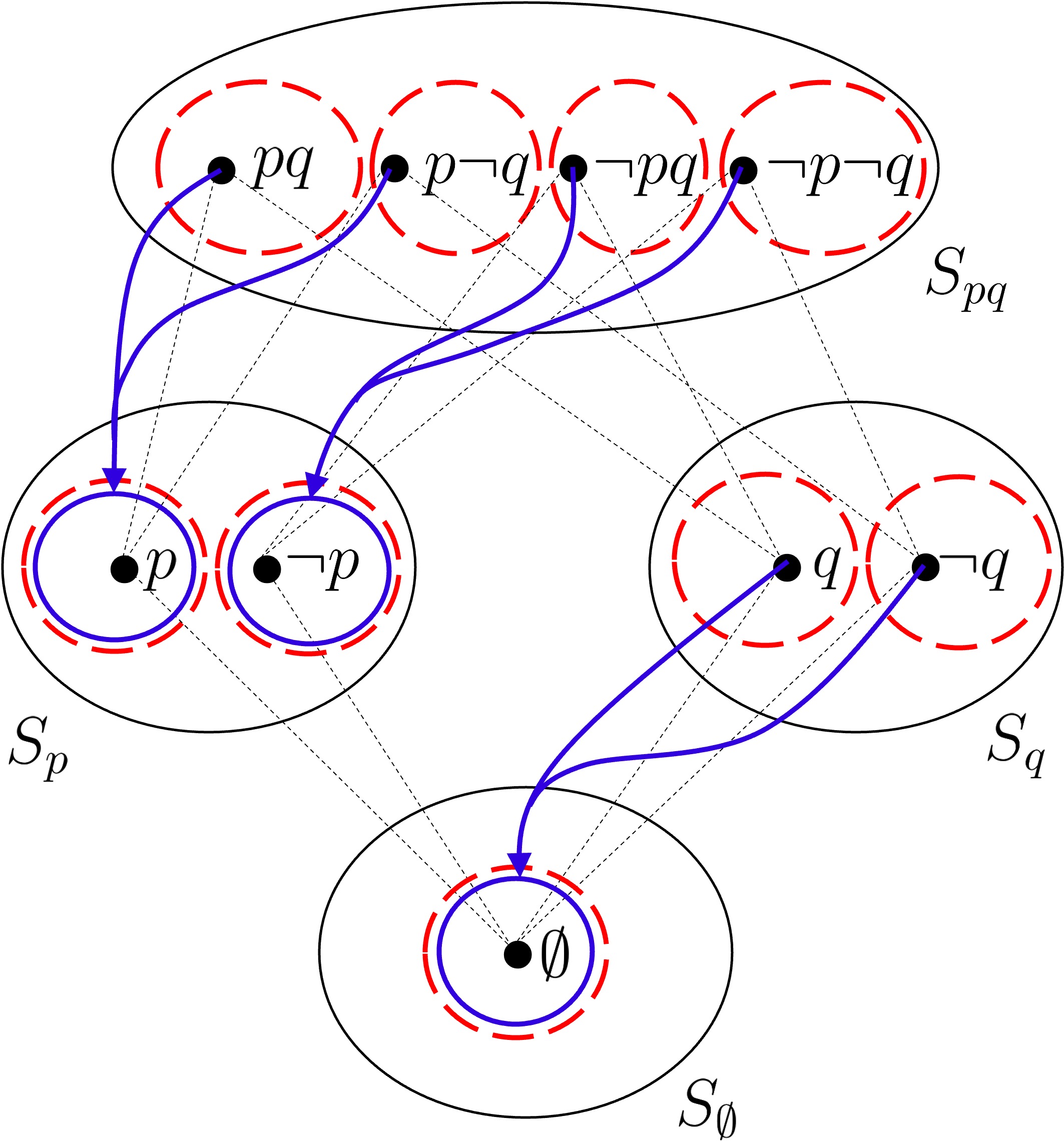}
\end{center}\underline{}
\end{figure}

Explicit Measurability says that explicit knowledge is measurable with respect to implicit knowledge. That is, the agent always implicitly knows her explicit knowledge. The converse, Implicit Measurability, is more subtle because of awareness. An individual may not explicitly know her implicit knowledge because she might be unaware of some events. However, the individual always explicitly knows her implicit knowledge modulo awareness. That is, she might implicitly know more at a higher awareness level than what she knows at her awareness level (like in the structure to the right in Figure~\ref{examples}) but at her awareness level, her implicit knowledge equals her explicit knowledge. The following lemma formalizes the last conclusion. The proof uses all properties of $\Pi_i$ and $\Lambda_i$ except Projections Preserve Knowledge of both $\Lambda_i$ and $\Pi_i$ and Projections Preserve Ignorance of $\Pi_i$.
	
\begin{lem}\label{coincides} For any individual $i \in I$, if $\omega' \in \Pi_i(\omega)$, then $\Lambda_i(\omega') = \Pi_i(\omega')$.
\end{lem}
		
\begin{lem}[Coherence]\label{coherence} For any individual $i \in I$, $\omega \in \Omega$, $\Lambda_i(\omega)_{S_{\Pi_i(\omega)}} = \Pi_i(\omega)$.
\end{lem}
	
Figure~\ref{examples} illustrates with two examples of how implicit knowledge can be ``fitted'' to explicit knowledge. Consider first the HMS model to the left. There are four spaces indexed by subsets of atomic formulae. Anticipating the semantics of HMS models introduced later, we describe and call states by their atomic formulae. The explicit possibility correspondence of the individual is indicated by the solid blue ovals and arrows. For instance, at state $pq$ she considers possible state $p$. That is, she is unaware of $q$ and knows $p$. Similarly, at state $\neg p q$ she is unaware of $q$ and knows $\neg p$. Her implicit possibility correspondence is given by the red dashed ovals. Note that in this complemented HMS model she does not implicitly know more than she does explicitly. Contrast this with the HMS model to the right. There, she implicitly knows $q$ for instance at state $pq$ (because her implicit possibility set at $pq$ is $\{pq\}$) although she is not aware of $q$ (because her explicit possibility set at $pq$ is on $S_{\{p\}}$). and hence does not explicitly know $q$. The figures demonstrate that the implicit possibility correspondence may be consistent with the explicit possibility correspondence in two different ways. It may model implicit knowledge that is finer than the explicit knowledge (like in the figure to the right) or implicit knowledge that is as coarse as the explicit knowledge but not coarser (like in the figure to the left). Note that a version of the models in Figure~\ref{examples} in which only $\{pq, p \neg q\}$ is in a red dashed oval while $\neg p q$ and $\neg p \neg q$ are in distinct circles in $S_{pq}$ is ruled out by Projections Preserve Implicit Knowledge. 

Given implicit possibility correspondences, we proceed with the definition of the implicit knowledge operators. 
	
\begin{defin}\label{definL} For any individual $i \in I$, the \emph{implicit knowledge operator} on events $E \in \Sigma$ is $$L_i(E) := \{\omega \in \Omega : \Lambda_i(\omega) \subseteq E\}$$ if there exists a state $\omega \in \Omega$ such that $\Lambda_i(\omega) \subseteq E$ and by $L_i(E) = \emptyset^{S(E)}$ otherwise. 
\end{defin}
	
The next observation follows immediately from the properties of the implicit possibility correspondence and the proof of Lemma~\ref{alles_events} in Heifetz, Meier, and Schipper (2006).
	
\begin{lem}\label{Levent} For any individual $i \in I$ and event $E \in \Sigma$, $L_i(E)$ is an $S(E)$-based event.
\end{lem}

Implicit knowledge satisfies all properties of ``partitional'' knowledge. 
	
\begin{prop}\label{S5_implicit} For any individual $i \in I$, $L_i$ satisfies for any $E, F \in \Sigma$ and $\{E_n\}_n \subseteq \Sigma$,
\begin{itemize}
\item[(i)] Necessitation: For $\Phi \subseteq \mathsf{At}$, $L_i(S_\Phi^\uparrow)= S_{\Phi}^\uparrow$,
			
\item[(ii)] Conjunction: $L_i\left(\bigcap_n E_n\right) = \bigcap_{n} L_i(E_n)$,
			
\item[(iii)] Monotonicity: $E \subseteq F$ implies $L_i(E) \subseteq L_i(F)$,
			
\item[(iv)] Truth: ${L}_i(E) \subseteq E$,
			
\item[(v)] Positive Introspection: $L_i(E) \subseteq L_i L_i(E)$,
			
\item[(vi)] Negative Introspection: $\neg L_i(E) \subseteq L_i \neg L_i(E)$.
\end{itemize}
\end{prop}
		
We observe that as in Fagin and Halpern (1988), explicit knowledge of an event equals implicit knowledge and awareness of that event.
\begin{prop}\label{prop: semantic equivalence of K - L and A} For any 
	$i \in I$ and event $E \in \Sigma$,
\begin{multicols}{2}
	\begin{enumerate}
\item\mbox{$K_i(E) =L_i(E) \cap A_i(E)$},
\item$U_i(E) = L_i(U_i(E))$,
\item$A_i(E) = L_i(A_i(E))$,
\item$A_i L_i(E) = A_i(E)$.
\end{enumerate}
\end{multicols}
\end{prop}
	
Properties 2. and 3. above mean that the individual implicitly knows her unawareness. This is in contrast to explicit knowledge since by KU introspection an individual can never explicitly know that she is unaware of an event. Property 4 says that an individual is aware of her implicit knowledge of an event if and only she is aware of the event. That is, the moment she can reason about an event, she can also reason about her implicit knowledge of the event. This is analogous to AK-Self-Reflection of explicit knowledge.

\section{From Implicit to Explicit\label{implicit-to-explicit}}
	
In the previous section, we showed how implicit knowledge can be derived from explicit knowledge. In this section, we go the other direction. We can devise a version of HMS model that features possibility correspondences capturing implicit knowledge and (non-syntactic) awareness functions as primitives, and then derive the possibility correspondence capturing explicit knowledge. 

\begin{defin} An~\emph{implicit knowledge-based HMS model} $\mathsf{M}^* = \langle I, \{S_{\Phi}\}_{\Phi \subseteq \mathsf{At}}, (r^{\Phi}_{\Psi})_{\Psi \subseteq \Phi \subseteq \mathsf{At}}, (\Lambda^*_i)_{i \in I}, (\alpha_i)_{i \in I}, \\ v \rangle$ consists~of 
\begin{itemize}
\item a non-empty set of individuals $I$,
\item a nonempty collection of nonempty disjoint state spaces $\{S_{\Phi}\}_{\Phi \subseteq \mathsf{At}}$ (as in Definition~\ref{HMS_model}),
\item projections $(r^{\Phi}_{\Psi})_{\Psi \subseteq \Phi \subseteq \mathsf{At}}$ (as in Definition~\ref{HMS_model}),
\item an implicit possibility correspondence $\Lambda_i^*: \Omega \longrightarrow 2^{\Omega} \setminus \{\emptyset\}$, for all $i\in I$,
\item an awareness function $\alpha_i: \Omega \longrightarrow \{S_\Phi\}_{\Phi \subseteq \mathsf{At}}$, for all $i\in I$,
\item a valuation function $v: \mathsf{At} \longrightarrow \Sigma$. 
\end{itemize}
\end{defin}

Like HMS models, implicit knowledge-based HMS models feature a projective lattice of state-spaces. However, instead of the explicit possibility correspondence, we now take the implicit possibility correspondences as a primitive. As before, we are interested in strong properties of knowledge associated with S5 because (1) these properties have been used for explicit knowledge in applications, and (2) we will require explicit knowledge to be consistent with implicit knowledge. As such, we are interested how the rich structure of S5 translates into properties of a derived explicit possibility correspondence. To that end, we require:  

\begin{ass} For each individual $i \in I$, the implicit possibility correspondence $\Lambda^*_i$ satisfies Reflexivity, Stationarity, and Projections Preserve Implicit Knowledge. 
\end{ass}

These properties were also satisfied by implicit possibility correspondences in the previous section.\footnote{\scriptsize Note again that Reflexivity and Stationarity implies Strong Confinement. In more general settings without Reflexivity or Stationarity, at least Strong Confinement would have to be imposed in $\Lambda_i^*$ for every $i \in I$.}

The second primitive of implicit knowledge-based HMS models is the awareness function $\alpha_i$ for every individual $i \in I$. We impose the following properties on $\alpha_i$: 

\begin{ass} For each individual $i \in I$, the awareness function $\alpha_i: \Omega \longrightarrow \{S_\Phi\}_{\Phi \subseteq \mathsf{At}}$ satisfies 
\begin{itemize}

\item[O.] Lack of Conception: If $\omega \in S_{\Phi}$, then $\alpha_i(\omega) \preceq S_{\Phi}$.

\item[I.] Awareness Measurability: If $\omega' \in \Lambda^*_i(\omega)$, then $\alpha_i(\omega') = \alpha_i(\omega)$.

\item[II.] If $\omega \in S_{\Phi}$ and $S_{\Psi} \preceq \alpha_i(\omega)$, then $\alpha_i(\omega_{\Psi}) = S_{\Psi}$.

\item[III.] If $\omega \in S_{\Phi}$ and $\alpha_i(\omega) \preceq S_\Psi \preceq S_\Phi$, then $\alpha_i(\omega_{\Psi}) = \alpha_i(\omega)$.

\item[IV.] If $\omega \in S_{\Phi}$ and $\Psi \subseteq \Phi$, then $\alpha_i(\omega) \succeq \alpha_i(\omega_{\Psi})$.
\end{itemize}
When $\alpha_i(\w) \in S$ for some $S \in \{S_{\Phi}\}_{\Phi \subseteq \mathsf{At}}$, we call $S$ the \emph{awareness level of $i$ at $\w$}.
\end{ass}
	
Property O. models one feature of Confinement of HMS models (see Assumption~\ref{assumptions_HMS}). Note that Confinement in HMS models has two features: First, it requires that the possibility set at a state is a subset of exactly one space. Second, it says that this space must be weakly less expressive than the space containing the state. Only this second last feature is captured by property O. The idea is that an individual may have lack of conception. Property I. is a measurability condition. Awareness is measurable with respect to implicit knowledge. The implication is that an agent implicitly knows her own awareness. Properties II. to IV. are consistency properties of awareness across the lattice. Projections preserve awareness as long as it is still expressible in the spaces. While property II. preserves awareness for corresponding states in spaces less expressive than the awareness level at a state, property III. preserves awareness for corresponding states in spaces more expressive than the awareness level at that state.

\begin{defin}\label{derived_explicit_possibility_correspondence} Given an implicit knowledge-based HMS model $\mathsf{M}^* = \langle I, \{S_{\Phi}\}_{\Phi \subseteq \mathsf{At}}, (r^{\Phi}_{\Psi})_{\Psi \subseteq \Phi \subseteq \mathsf{At}}, (\Lambda^*_i)_{i \in I}, \\ (\alpha_i)_{i \in I}, v \rangle$, define the explicit possibility correspondence $\Pi^*_i: \Omega \longrightarrow 2^{\Omega}$ by, for all $\omega \in \Omega$ and $\Phi \subseteq \mathsf{At}$, $\Pi^*_i(\omega_{\Phi}) := \Lambda^*_i(\omega)_{\alpha_i(\omega_{\Phi})}$. We call \ $\overline{\mathsf{M}}^* = \langle I, \{S_{\Phi}\}_{\Phi \subseteq \mathsf{At}}, (r^{\Phi}_{\Psi})_{\Psi \subseteq \Phi \subseteq \mathsf{At}}, (\Lambda^*_i)_{i \in I}, (\Pi^*_i)_{i \in I}, (\alpha_i)_{i \in I}, v \rangle$ the \emph{complemented implicit knowledge-based HMS model}. 
\end{defin}
	
The defining condition for the explicit possibility correspondence in implicit knowledge-based HMS models is a slight strengthening of Coherence derived from the explicit and implicit measurability in Lemma~\ref{coherence}. Here we take it as the primitive to connect explicit knowledge to implicit knowledge.
	
The following observations are immediate:
\begin{lem}\label{immediate} For all $\omega \in \Omega$,
\begin{itemize}
\item[A.] $\Pi^*_i(\omega) = \Lambda^*_i(\omega)_{\alpha_i(\omega)}$,
\item[B.] $\Pi^*_i(\omega_{\Phi}) = \Lambda^*_i(\omega)_{\Phi}$ for all $\Phi \subseteq \mathsf{At}$ with $S_\Phi \preceq \alpha_i(\omega)$,
\item[C.] $\Pi^*_i(\omega_{\Phi}) = \Lambda^*_i(\omega)_{\alpha_i(\omega)}$ for all $\Phi \subseteq \mathsf{At}$ with $S_{\omega} \succeq S_\Phi \succeq \alpha_i(\omega)$.
\end{itemize}
\end{lem}
	
The following lemma records properties of the derived explicit possibility correspondence. It shows that it satisfies the properties of the explicit possibility correspondence of HMS models. 

\begin{lem} For any individual $i \in I$, if $\alpha_i$ satisfies O., I., II., III., and IV., then $\Pi^*_i$ satisfies Confinement, Generalised Reflexivity, Stationarity, Projections Preserve Ignorance, Projections Preserve Knowledge.
\end{lem}
	
We conclude that the derived explicit possibility correspondence $\Pi^*_i$ is a possibility correspondence as in Heifetz, Meier, and Schipper (2006, 2008), i.e., satisfies Assumption~\ref{assumptions_HMS}. To show that the connection between the derived explicit possibility correspondence and the implicit possibility correspondence is as in the complemented HMS model of the prior section, we note the following lemma. 

\begin{lem}\label{measurability_return} For any individual $i \in I$, $\Lambda_i^*$ and $\Pi_i^*$ jointly satisfy explicit and implicit measurability. 
\end{lem}

The above lemmata imply the following: 

\begin{cor}\label{cHMS_derived} For any implicit knowledge-based HMS model $\mathsf{M}^* = \langle I, \{S_{\Phi}\}_{\Phi \subseteq \mathsf{At}}, (r^{\Phi}_{\Psi})_{\Psi \subseteq \Phi \subseteq \mathsf{At}}, (\Lambda^*_i)_{i \in I}, \\ (\alpha_i)_{i \in I}, v \rangle$ with derived explicit possibility correspondences $(\Pi^*_i)_{i \in I}$ we have that $\overline{\mathsf{M}} = \langle I, \{S_{\Phi}\}_{\Phi \subseteq \mathsf{At}}, \\ (r^{\Phi}_{\Psi})_{\Psi \subseteq \Phi \subseteq \mathsf{At}}, (\Lambda^*_i)_{i \in I}, (\Pi^*_i)_{i \in I}, v \rangle$ is a complemented HMS model and $\mathsf{M} = \langle I, \{S_{\Phi}\}_{\Phi \subseteq \mathsf{At}}, (r^{\Phi}_{\Psi})_{\Psi \subseteq \Phi \subseteq \mathsf{At}}, \\ (\Pi^*_i)_{i \in I}, v \rangle$ is a HMS model. 
\end{cor}

The awareness function can be directly used to define an awareness operator on events. 

\begin{defin} For each individual $i \in I$, define an awareness operator on events by for all $E \in \Sigma$, $A^*_i(E) := \{\omega \in \Omega : \alpha_i(\omega) \succeq S(E)\}$ if there is a state $\omega \in \Omega$ such that $\alpha_i(\omega) \succeq S(E)$ and by $A^*_i(E) = \emptyset^{S(E)}$ otherwise.
\end{defin} 

Similarly, for each individual $i \in I$, we can use the possibility correspondence $\Lambda^*_i$ to define an implicit knowledge operator $L^*_i$ as in Definition~\ref{definL}. Finally, let $K_i$ be the explicit knowledge operator and $A_i$ be the awareness operator defined from the derived explicit possibility correspondence $\Pi^*_i$ as in Definitions~\ref{definK} and~\ref{definA}, respectively.
 
The following proposition shows that awareness defined with the awareness function is equivalent to awareness defined with the derived explicit possibility correspondence. It also shows that explicit knowledge defined from the derived explicit possibility correspondence is equivalent to implicit knowledge and awareness. 

\setlength{\columnsep}{-20pt}
\begin{prop}\label{otherwayaround} For every 
$i \in I$ and any event $E \in \Sigma$,
\begin{multicols}{1}
\begin{enumerate}
\item $A^*_i(E) = A_i(E)$
\item\mbox{$K_i(E) = L^*_i(E) \cap A^*_i(E)$}
\end{enumerate}
\end{multicols}
\end{prop}
The last two sections show an interdefinability of explicit and implicit knowledge in HMS models. Implicit knowledge can be defined in terms of explicit knowledge and vice versa. We can use either the explicit possibility correspondences as primitive or the implicit possibility correspondence together with the awareness function. Implicit knowledge-based HMS models are arguably closer to FH models than HMS models. We will use them to build a bridge between HMS and FH models.

\section{Category of FH Models}
	
In this section, we introduce FH models and bounded morphisms, a notion of structure preserving maps between FH models, and use these to form a category with FH models as objects and bounded morphisms as morphisms. 

The semantics of FH models is not syntax-free since each agent's awareness function assigns to each state a set of formulae. Thus, we first introduce the formal language featuring implicit knowledge, awareness, and explicit knowledge. With $i \in I$ and $p \in \mathsf{At}$, define the language $\mathcal{L}_{\mathsf{At}}$ by
$$\varphi::=\top \mid p \mid \neg\varphi \mid \varphi \wedge \psi \mid \ell_i \varphi \mid a_i \varphi \mid k_i \varphi$$
	
Let $\mathsf{At}(\varphi) := \{ p \in \mathsf{At} \colon p \text{ is a subformula of } \varphi\}$, for any $\varphi \in \mathcal{L}_{\mathsf{At}}$, be the set of atomic formulae that are contained in $\varphi$, and let $\mathcal{L}_\Phi := \{\varphi \in \mathcal{L}_{\mathsf{At}}: \mathsf{At}(\varphi) \subseteq \Phi \}$ be the sublanguage of $\mathcal{L}_{\mathsf{At}}$ built on propositions $p$ in $\Phi \subseteq \mathsf{At}$. 
	
The formula $\ell_i \varphi$ reads ``agent $i$ implicitly knows $\varphi$",  $a_i \varphi$ reads ``$i$ is aware of $\varphi$", and $k_i \varphi$ reads ``$i$ explicitly knows $\varphi$". Fagin and Halpern (1988) define explicit knowledge as the conjunction of implicit knowledge and awareness, namely $k_{i} \varphi = a_i \varphi \wedge \ell_i \varphi$, for $\varphi \in \mathcal{L}_{\mathsf{At}}$.

\begin{defin}\label{def:FH model} For any $\Phi \subseteq \mathsf{At}$, a \emph{FH model} $\mathsf{K}_{\Phi} = \langle I, W_{\Phi}, (R_{\Phi, i})_{i \in I}, (\mathcal{A}_{\Phi,i})_{i \in I}, V_{\Phi}\rangle$ for $\Phi$ consists of 
\begin{itemize}
\item a non-empty set of individuals $I$,
\item a non-empty set of states $W_{\Phi}$,
\item an accessibility relation $R_{\Phi, i} \subseteq W_{\Phi} \times W_{\Phi}$, for all $i\in I$,
\item an awareness function $\mathcal{A}_{\Phi, i}: W_{\Phi} \longrightarrow 2^{\mathcal{L}_{\Phi}}$, for all $i\in I$, assigning to each state $w \in W_{\Phi}$ a set of formulas $\mathcal{A}_{\Phi, i}(w) \subseteq \mathcal{L}_{\Phi}$. The set $\mathcal{A}_{\Phi, i}(w)$ is called the \emph{awareness set of $i$ at $w$}. 
\item a valuation function $V_{\Phi}: \Phi \longrightarrow 2^{W_{\Phi}}$. 
\end{itemize}
\end{defin} 

\begin{ass} We require that the FH model $K_{\Phi}$ is \emph{propositionally determined}, i.e., for every $i \in I$, the awareness functions satisfy
\begin{itemize}
\item[] \emph{Awareness is Generated by Primitive Propositions:} For all $\varphi \in \mathcal{L}_{\Phi}$, $\varphi \in \mathcal{A}_{\Phi, i}(w)$ if and only if for all $p \in \mathsf{At}(\varphi)$, $p \in \mathcal{A}_{\Phi, i}(w)$.
\item[] \emph{Agents Know What They Are Aware of:} $(w, t) \in R_{\Phi, i}$ implies $\mathcal{A}_{\Phi, i}(w) = \mathcal{A}_{\Phi, i}(t)$.
\end{itemize}
We also require that the FH model $\mathsf{K}_{\Phi}$ is \emph{partitional}, that is, $R_{\Phi, i}$ is an equivalence relation, i.e., satisfies reflexivity, transitivity, and Euclideaness. 
\end{ass} Throughout the paper, we focus on partitional and propositionally determined FH models because these models capture the notion of awareness and knowledge used in most applications so far and it is also the notion of awareness used in HMS models. We are interested in how this rich structure maps between FH models as well as between FH and HMS models. 
 
\begin{defin} For any $\Psi \subseteq \Phi \subseteq \mathsf{At}$ and FH models $\mathsf{K}_{\Phi} = \langle I, W_{\Phi}, (R_{\Phi, i})_{i \in I}, (\mathcal{A}_{\Phi,i})_{i \in I}, V_{\Phi}\rangle$ and $\mathsf{K}_{\Psi} = \langle I, W_{\Psi}, (R_{\Psi, i})_{i \in I}, (\mathcal{A}_{\Psi,i})_{i \in I}, V_{\Psi}\rangle$, the mapping $f^{\Phi}_{\Psi}: \mathsf{K}_{\Phi} \longrightarrow \mathsf{K}_{\Psi}$ is a \emph{surjective bounded morphism} if for every $i \in I$ and $w\in W_\Phi$
\begin{itemize}
\item \emph{Surjectivity:} $f^{\Phi}_{\Psi}: W_{\Phi} \longrightarrow W_{\Psi}$ is a surjection,

\item \emph{Atomic harmony:} for every $p \in \Psi$, $w \in V_{\Phi}(p)$ if and only if $f^{\Phi}_{\Psi}(w) \in V_{\Psi}(p)$,

\item \emph{Awareness consistency:} $\mathcal{A}_{\Phi, i}(w) \cap \mathcal{L}_{\Psi} = \mathcal{A}_{\Psi, i}(f^{\Phi}_{\Psi}(w))$

\item \emph{Homomorphism:} $f^{\Phi}_{\Psi}$ is a homomorphism w.r.t. $R_{\Phi, i}$, i.e., if $(w, t) \in R_{\Phi, i}$, then $(f^{\Phi}_{\Psi}(w),f^{\Phi}_{\Psi}(t))\in R_{\Psi,i}$, 
\item \emph{Back:}  
if $(f^{\Phi}_{\Psi}(w),t') \in R_{\Psi, i}$, then there is a state $t \in W_{\Phi}$ such that $f^{\Phi}_{\Psi}(t) = t'$ and $(w, t) \in R_{\Phi, i}$.
\end{itemize}
\end{defin}

This is the standard notion of bounded morphism (also called $p$-morphism) (see for instance, Blackburn, de Rijke, and Venema, 2001, pp. 59--62) except for the additional property of Awareness Consistency. In our context, the bounded morphism literally bounds the language over which FH models are defined. We can now consider collections of FH models and bounded morphisms between them:
 
\begin{defin}\label{category} Given the FH model $\mathsf{K}_{\mathsf{At}}$, the \emph{category of FH models} $\mathcal{C}(\mathsf{K}_{\mathsf{At}}) = \langle (\mathsf{K}_{\Phi})_{\Phi \subseteq \mathsf{At}}, (f^{\Phi}_{\Psi})_{\Psi \subseteq \Phi \subseteq \mathsf{At}} \rangle$ consists of 
\begin{itemize}
\item a collection of FH models $\mathsf{K}_{\Phi}$, one for each $\Phi \subseteq \mathsf{At}$,
\item for any $\Phi, \Psi \subseteq \mathsf{At}$ with $\Psi \subseteq \Phi$, there is one surjective bounded morphism $f^{\Phi}_{\Psi}$, such that 

\smallskip
-- for any $\Phi \subseteq \mathsf{At}$, $f^{\Phi}_{\Phi}$ is the identity,
 
\smallskip
-- for any $\Upsilon, \Phi, \Psi \subseteq \mathsf{At}$ with $\Upsilon \subseteq \Psi \subseteq \Phi$, $f^{\Phi}_{\Upsilon} = f^{\Psi}_{\Upsilon} \circ f^{\Phi}_{\Psi}$. 
\end{itemize}
 
\end{defin}

Our terminology is not arbitrary. The category of FH models is indeed a category in the sense of category theory. It has an initial object, the most expressive FH model $\mathsf{K}_{\mathsf{At}}$, as well as a terminal object, the FH model $\mathsf{K}_{\emptyset}$. 

Since the category of FH models is defined with bounded morphisms, it suggests that all FH models in the category are in some sense epistemically equivalent. Indeed, we interpret each category of FH models literally as the category of FH models that vary with the language but are otherwise modally equivalent. That is, for any $\Psi \subseteq \Phi \subseteq \mathsf{At}$, modal satisfaction for $\mathsf{K}_{\Psi}$ is as for $\mathsf{K}_{\Phi}$ w.r.t. formulae in $\mathcal{L}_{\Psi}$ (see Lemma~\ref{boring_modal_equivalence} below). We interpret this as follows: When a modeler represents a context with a FH model $\mathsf{K}_{\mathsf{At}}$, an agent $i$ at state $w \in W_{\mathsf{At}}$ can be thought of representing it with the FH model $\mathsf{K}_{\mathsf{At}(\mathcal{A}_{\mathsf{At}, i}(w))}$. And this agent $i$ considers it possible at $w$ that at $t$ with $(f^{\mathsf{At}}_{\mathsf{At}(\mathcal{A}_{\mathsf{At}, i}(w))}(w), t) \in R_{\mathsf{At}(\mathcal{A}_{\mathsf{At}, i}(w)), i}$ agent $j$ represents the situation with the FH model $\mathsf{K}_{\mathsf{At}(\mathcal{A}_{\mathsf{At}, j}(t))}$, etc. These models can all be seen as equivalent except for the language of which they are defined. With this construction, we do not just endow agents with a formal language to reason about their context but we also allow them to analyze their context with semantic devices like logicians do. This is relevant because in many multi-agent contexts of game theory, the analysis proceeds using semantic devices like state spaces etc. rather than at the level of syntax. For instance, in a principal-agent problem, the principal may want to use a FH model augmented by actions and utility functions to analyze optimal contract design realizing that the (unaware) agent may also use a less expressive but otherwise equivalent FH model to analyze how to optimally interact with the principal. 

To make the equivalence between models in the category of FH models precise, we need to introduce the semantics of FH models. 
\begin{defin} For any $\Phi \subseteq \mathsf{At}$, FH model $\mathsf{K}_{\Phi} = \langle I, W_{\Phi}, (R_{\Phi, i})_{i \in I}, (\mathcal{A}_{\Phi,i})_{i \in I}, V_{\Phi}\rangle$, and $\omega \in W_{\Phi}$, satisfaction of formulae in $\mathcal{L}_{\Phi}$ is given by the following clauses:
\begin{center}
	\begin{tabular}{lllcclll}
		    $\mathsf{K}_{\Phi}, w \Vdash \top$ &  & for all $w\in W_\Phi$; & & & $\mathsf{K}_{\Phi}, w \Vdash \varphi\wedge \psi$ &  iff & $\mathsf{K}_{\Phi}, w \Vdash \varphi$ and $\mathsf{K}_{\Phi}, w \Vdash \psi$; \tabularnewline
			$\mathsf{K}_{\Phi}, w \Vdash p$ &  iff & $w \in V_{\Phi}(p)$; & & & $\mathsf{K}_{\Phi},w \Vdash a_i\varphi$ & iff & $\varphi\in\mathcal{A}_{\Phi, i}(w)$; \tabularnewline
			$\mathsf{K}_{\Phi}, w \Vdash \neg \varphi$ &  iff & $\mathsf{K}_{\Phi}, w \not\Vdash \varphi$; & & & $\mathsf{K}_{\Phi}, w \Vdash \ell_i \varphi$ & iff & $\mathsf{K}_{\Phi}, t\Vdash \varphi$ for all $(w, t) \in R_{\Phi, i}$.\tabularnewline		
	\end{tabular}
\end{center}
\end{defin}
	
From this semantics and the syntactic definition $k_i\varphi := \ell_i \varphi \wedge a_i \varphi$, it follows that $\mathsf{K}_{\Phi}, w \Vdash k_i \varphi$ if and only if for all $t$ s.t. $(w, t) \in R_{\Phi, i}$, $\mathsf{K}_{\Phi},t \Vdash \varphi$ and $\varphi \in \mathcal{A}_{\Phi, i}(w)$.

The category of FH models forms a complete lattice induced by set inclusion on sets of atomic formulae with the initial object being the join of the lattice and the terminal object being the meet of the lattice. We now show that it gives rise to a complete lattice when ordered using the (directed) bounded morphism or, epistemically more relevant, when ordered by modal equivalence relative to sublanguages. 

\begin{prop}\label{epistemic_equivalence} Given a category of FH models, $\langle (\mathsf{K}_{\Phi})_{\Phi \subseteq \mathsf{At}}, (f^{\Phi}_{\Psi})_{\Psi \subseteq \Phi \subseteq \mathsf{At}} \rangle$, modal equivalence relative to sublanguages forms a complete lattice of FH models in the category as follows: For any nonempty set of subsets of atomic formulae $\mathcal{F} \subseteq 2^{\mathsf{At}}$,   
\begin{itemize}
\item[(i)] $\mathsf{K}_{\bigcup_{\Phi \in \mathcal{F} } \Phi}$ is modally equivalent to $\mathsf{K}_{\Psi}$ w.r.t. $\mathcal{L}_{\Psi}$ for every $\Psi \in \mathcal{F} $, i.e., for any $w \in W_{\bigcup_{\Phi \in \mathcal{F} } \Phi}$, $\varphi \in\mathcal{L}_\Psi$, $\mathsf{K}_{\bigcup_{\Phi \in \mathcal{F} } \Phi}, w \Vdash \varphi$ iff $\mathsf{K}_\Psi, f^{\bigcup_{\Phi \in \mathcal{F} } \Phi}_{\Psi}(w) \Vdash \varphi$, and 
\item[(ii)] $\mathsf{K}_{\bigcap_{\Phi \in \mathcal{F}} \Phi}$ is modally equivalent to $\mathsf{K}_{\Psi}$ w.r.t. $\mathcal{L}_{\bigcap_{\Phi \in \mathcal{F}}}$ for every $\Psi \in \mathcal{F}$, i.e., for any $w \in W_{\Psi}$, $\varphi \in \mathcal{L}_{\bigcap_{\Phi \in \mathcal{F}} \Phi}$, $\mathsf{K}_\Psi, w \Vdash \varphi$ iff $\mathsf{K}_{\bigcap_{\Phi \in \mathcal{F}} \Phi}, f^{\Psi}_{\bigcap_{\Phi \in \mathcal{F} } \Phi}(w) \Vdash \varphi$.
\end{itemize}
\end{prop}

The proof of the proposition uses of the following lemma: 

\begin{lem}\label{boring_modal_equivalence} For any $\Psi, \Phi \subseteq \mathsf{At}$ with $\Psi \subseteq \Phi$, all $w \in W_{\Phi}$, and all $\varphi \in \mathcal{L}_{\Psi}$, $\mathsf{K}_{\Phi}, w \Vdash \varphi$ if and only if $\mathsf{K}_{\Psi}, f^{\Phi}_{\Psi}(w) \Vdash \varphi.$
\end{lem}

Note that for a collection of FH models $\{\mathsf{K}_{\Psi}\}_{\Psi \in \mathcal{F}}$, the ``join'' and ``meet'' FH models are $\mathsf{K}_{\bigcup_{\Phi \in \mathcal{F} } \Phi}$ and $\mathsf{K}_{\bigcap_{\Phi \in \mathcal{F}} \Phi}$, respectively. So for any collection of FH models, Proposition~\ref{epistemic_equivalence} shows modal equivalence between any FH model in the collection and its join and meet models, respectively.

Our notion of bounded morphism  is inspired by bisimulation of FH models introduced by van Ditmarsch et al. (2018). Clearly, the surjective bounded morphism is a bisimulation. Here we discuss some differences and similarities. While bisimulation more generally is a relation between models without a particular direction, the bounded morphism has a natural direction from the more expressive FH model to the less expressive FH model. Further, it is a function on $W_{\Phi}$. That is, it maps \emph{every} state in $W_{\Phi}$ to a state in $W_{\Psi}$ with $\Psi \subseteq \Phi$. Moreover, surjectivity is a property that is straightforward to define for functions. Finally, bounded morphisms easily compose and almost naturally lead to the notion of category of FH models although we do not really make much use here of the machinery of category theory. For all these reasons, we use the notion of bounded morphism. Van Ditmarsch et al. (2018) introduced two notions of bisimulation for FH models, standard bisimulation and awareness bisimulation. Like our bounded morphism, both of their notions of bisimulation also depend on a subset of atomic formulae for FH models. The clauses Atomic harmony, Awareness consistency, Homomorphism, and Back have counterparts in their notions of bisimulations. Our notion of bounded morphism is closer to what they call standard bisimulation because our Homomorphism and Back clauses do not involve the awareness function. Although van Ditmarsch et al. (2018, p. 63) mention the projective lattice structure of Heifetz, Meier, and Schipper (2006, 2008) as a motivation for their notion of awareness bisimulation, we believe it is particularly useful for their notion of speculative knowledge. Their notions of bisimulations do not require surjectivity although when considering maximal bisimulations, they must be surjective since they yield quotient models. Compositions of maximal bisimulation commute like we require our bounded morphism to do in the category of FH models but bisimulations that are not maximal do not necessarily commute. Moreover, maximal bisimulations yield necessarily contractions that eliminate redundancies. We are unsure whether it is necessarily a natural property when we interpret categories of FH models as collections of subjective views of agents. An agent may not realize or may not be bothered by redundancies and use an FH model with redundancies to analyze her situation. That is, differences in awareness and redundancies are orthogonal to each other and reduction in awareness does not necessitate elimination of redundancies.

\section{Transformations}

\subsection{From FH Models to HMS Models}\label{sec: u-transf}

We can use the tools of the prior sections to define a transformation of a FH model into a HMS model. The transformation works as follows: For any FH model $\mathsf{K}_{\mathsf{At}}$ for $\mathsf{At}$, consider the category of FM models $\langle (\mathsf{K}_{\Phi})_{\Phi \subseteq \mathsf{At}}, (f^{\Phi}_{\Psi})_{\Psi \subseteq \Phi \subseteq \mathsf{At}} \rangle$. This category is transformed into an implicit knowledge-based HMS model. We then derive the explicit possibility correspondences and add them to the implicit knowledge-based HMS model, obtaining a complemented implicit knowledge-based HMS model. In the next step, we erase the awareness functions and get a complemented HMS model. The core step is to transform a category of FH models into an implicit knowledge-based HMS model. This is defined next.  

\begin{defin}\label{T_transform} For any category of FH models $\mathcal{C}(\mathsf{K}_{\mathsf{At}}) = \langle (\mathsf{K}_{\Phi})_{\Phi \subseteq \mathsf{At}}, (f^{\Phi}_{\Psi})_{\Psi \subseteq \Phi \subseteq \mathsf{At}} \rangle$, the \emph{$T$-transform} of $\mathcal{C}(\mathsf{K}_{\mathsf{At}})$ is the implicit knowledge-based HMS model $T(\mathcal{C}(\mathsf{K}_{\mathsf{At}})) = \langle I, \{S_{\Phi}\}_{\Phi \subseteq \mathsf{At}}, (r^{\Phi}_{\Psi})_{\Psi \subseteq \Phi \subseteq \mathsf{At}}, (\Lambda^*_i)_{i \in I}, \\ (\alpha_i)_{i \in I}, v \rangle$ defined by:
\begin{itemize}
\item $S_{\Phi} = W_{\Phi}$ for all $\Phi \subseteq \mathsf{At}$, where $W_\Phi$ is the state space of the FH model $\mathsf{K}_{\Phi}$ of the category $\mathcal{C}(\mathsf{K}_{\mathsf{At}})$. Denote $\Omega = \bigcup_{\Phi \subseteq \mathsf{A}} S_{\Phi}$.
\item $r^{\Phi}_{\Psi} = f^{\Phi}_{\Psi}$ for any $\Phi, \Psi \subseteq \mathsf{At}$ with $\Psi \subseteq \Phi$, where $f^{\Phi}_{\Psi}$ is the surjective bounded morphism of the category $\mathcal{C}(\mathsf{K}_{\mathsf{At}})$. 
\item $\Lambda^*_i: \Omega \longrightarrow 2^\Omega$ such that $\Phi \subseteq \mathsf{At}$ and $w \in S_{\Phi}$, $w' \in \Lambda^*_i(w)$ if and only if $(w, w') \in R_{\Phi,i}$, for any $i \in I$,
\item $\alpha_i: \Omega \longrightarrow \{S_{\Phi}\}_{\Phi \subseteq \mathsf{At}}$ such that for all $\Psi \subseteq \mathsf{At}$ and $w \in S_{\Psi}$, $\alpha_i(w) = S_{\Upsilon}$ if and only if $\mathsf{At}(\mathcal{A}_{\Psi, i}(w)) = \Upsilon$, for any $i \in I$, 
\item  $v(p) = \bigcup_{\Phi \subseteq \mathsf{At}} V_{\Phi}(p)$, for any $p \in \mathsf{At}$.
\end{itemize}
\end{defin}

The $T$-transform indeed transforms any category of $FH$ models into an implicit knowledge-based HMS model. 

\begin{prop}\label{prop: T transform} For any category of FH models $\mathcal{C}(\mathsf{K}_{\mathsf{At}})$, the $T$-transform $T(\mathcal{C}(\mathsf{K}_{\mathsf{At}}))$ is an implicit knowledge-based HMS model.
\end{prop}

We have all ingredients to define the transformation of FH models into complemented HMS models. 

\begin{defin} For any FH model $\mathsf{K}_{\mathsf{At}}$, the \emph{HMS transform} $HMS(\mathsf{K}_{\mathsf{At}}) = \langle I, \{S_{\Phi}\}_{\Phi \subseteq \mathsf{At}}, (r^{\Phi}_{\Psi})_{\Psi \subseteq \Phi \subseteq \mathsf{At}}, \\ (\Lambda^*_i)_{i \in I}, (\Pi^*_i)_{i \in I}, v \rangle$ is defined by the following steps:
\begin{enumerate}
\item Form the category of FH models $\mathcal{C}(\mathsf{K}_{\mathsf{At}})$ (Definition~\ref{category}).
\item Apply the $T$-transform to $\mathcal{C}(\mathsf{K}_{\mathsf{At}})$ to obtain the implicit knowledge-based HMS model $T(\mathcal{C}(\mathsf{K}_{\mathsf{At}})) = \langle I, \{S_{\Phi}\}_{\Phi \subseteq \mathsf{At}}, (r^{\Phi}_{\Psi})_{\Psi \subseteq \Phi \subseteq \mathsf{At}}, (\Lambda^*_i)_{i \in I}, (\alpha_i)_{i \in I}, v \rangle$ (Defin.~\ref{T_transform}).
\item Form the complemented implicit knowledge-based HMS model $\overline{T(\mathcal{C}(\mathsf{K}_{\mathsf{At}}))} = \langle I, \{S_{\Phi}\}_{\Phi \subseteq \mathsf{At}}, \\ (r^{\Phi}_{\Psi})_{\Psi \subseteq \Phi \subseteq \mathsf{At}}, (\Lambda^*_i)_{i \in I}, (\Pi^*_i)_{i \in I}, (\alpha_i)_{i \in I}, v \rangle$ by deriving the explicit possibility correspondences $(\Pi^*_i)_{i \in I}$ (Definition~\ref{derived_explicit_possibility_correspondence}). 
\item Erase the awareness functions $(\alpha_i)_{i \in I}$ from the complemented implicit knowledge-based HMS model $\overline{T(\mathcal{C}(\mathsf{K}_{\mathsf{At}}))}$ to obtain the complemented HMS model $\langle I, \{S_{\Phi}\}_{\Phi \subseteq \mathsf{At}}, (r^{\Phi}_{\Psi})_{\Psi \subseteq \Phi \subseteq \mathsf{At}}, (\Lambda^*_i)_{i \in I}, \\ (\Pi^*_i)_{i \in I}, v \rangle$. 
\end{enumerate}
\end{defin}

\begin{cor} \label{cor: transform2} For any FH model $\mathsf{K}_{\mathsf{At}}$, its $HMS(\mathsf{K}_{\mathsf{At}})$ is a complemented HMS model.
\end{cor}

\subsection{From HMS Models to FH Models}\label{sec: fh-transf}

To transform a complemented HMS model into a FH model we simply need to consider the upmost space of the lattice of spaces of the HMS model, copy  the domain, define accessibility relations from implicit possibility correspondences, and the valuation function, and, for every state $\w \in S_{At}$, construct the awareness set at $\w$ by collecting all the formulas that contain the atoms defined in the space where $\Pi_i(\w)$ lies. 
	
\begin{defin} For any complemented HMS model $\overline{\mathsf{M}} = \langle I, \{S_{\Phi}\}_{\Phi \subseteq \mathsf{At}}, (r^{\Phi}_{\Psi})_{\Psi \subseteq \Phi \subseteq \mathsf{At}}, (\Lambda_i)_{i \in I}, (\Pi_i)_{i \in I}, v \rangle$, the  \emph{$FH$-transform} $FH(\overline{\mathsf{M}}) = \langle I, W_{\mathsf{At}}, (R_{\mathsf{At}, i})_{i \in I}, (\mathcal{A}_{\mathsf{At},i})_{i \in I}, V_{\mathsf{At}}\rangle$ is defined by 
\begin{itemize}
\item $W_{\mathsf{At}} = S_{\mathsf{At}},$
\item $R_{\mathsf{At}, i} \subseteq W_{\mathsf{At}} \times W_{\mathsf{At}}$ is such that $(\w, \w') \in R_{\mathsf{At}, i}$ if and only if $\w' \in \Lambda_i(\w)$, for all $i \in I$, 
\item $\mathcal{A}_{\mathsf{At}, i}: W_{\mathsf{At}} \longrightarrow 2^{\mathcal{L}_{\mathsf{At}}}$ is such that $\mathcal{A}_{\mathsf{At}, i} (\w) = \mathcal{L}_{\Phi}$ for $\Phi \subseteq \mathsf{At}$ with $\Pi_i(\w) \subseteq S_{\Phi}$, for all $i \in I$,
\item $V_{\mathsf{At}}: \mathsf{At} \longrightarrow 2^{W_{\mathsf{At}}}$ is such that $V_{\mathsf{At}}(p) = v(p) \cap S_{\mathsf{At}}$, for every $p \in \mathsf{At}$.
\end{itemize}
\end{defin}

The FH transform indeed transforms any complemented HMS model into a FH model. 

\begin{prop}\label{prop: transform1} For every complemented HMS model $\overline{\mathsf{M}}$, $FH(\overline{\mathsf{M}})$ is a FH model for $\mathsf{At}$.
\end{prop}

\subsection{Equivalence of HMS and FH Models\label{subsec: equivalence LKA}}

Before we can prove an equivalence of HMS and FH models, we need to introduce the semantics of complemented HMS models. 

\begin{defin}\label{def: satisfiability of L} Let $\overline{\mathsf{M}} = \langle I, \{S_{\Phi}\}_{\Phi \subseteq \mathsf{At}}, (r^{\Phi}_{\Psi})_{\Psi \subseteq \Phi \subseteq \mathsf{At}}, (\Lambda_i)_{i \in I}, (\Pi_i)_{i \in I}, v \rangle$ be a complemented HMS model and let $\w \in \Omega$. Satisfaction of $\mathcal{L}_{\mathsf{At}}$ formulas in $\overline{\mathsf{M}}$ is given by \  $\overline{\mathsf{M}},\w \vDash \top$ for all $\w \in \Omega$;
\begin{center}
	\begin{tabular}{lllcclll}
	$\overline{\mathsf{M}},\w \vDash p$ & $ \text{ iff }$ & $\w \in v(p)$; & & & $\overline{\mathsf{M}}, \w \vDash a_i \varphi$ & $\text{ iff } $& $S_{\Pi_i(\w)} \succeq S([\varphi])$; \tabularnewline
	$\overline{\mathsf{M}}, \w \vDash \neg \varphi$ & $\text{ iff }$ & $\w \in \neg [\varphi]$; & & & $\overline{\mathsf{M}},\w \vDash \ell_i\varphi$ & $ \text{ iff } $& $\Lambda_i(\w)\subseteq [\varphi]$; \tabularnewline
	$\overline{\mathsf{M}}, \w \vDash \varphi \wedge \psi $& $\text{ iff }$ & $ \w \in [\varphi] \cap [\psi]$; & & & $\overline{\mathsf{M}},\w \vDash k_i \varphi$ & $ \text{ iff }$ & $\Pi_i(\w) \subseteq [\varphi]$;\tabularnewline
\end{tabular}
\end{center}
\noindent where $[\varphi] := \{\w' \in \Omega : \overline{\mathsf{M}}, \w' \vDash \varphi\}$ for all $\varphi \in \mathcal{L}_{\mathsf{At}}$.
\end{defin}

In HMS models, formulae may have undefined truth value since formulae may not be even defined in every state. The same happens in FH models of a category of FH models. For instance, the truth value of $p$ is not defined for all FH models $\mathsf{K}_{\Phi}$ with $\Phi \not\ni p$. We will return to this issue later when we prove soundness and completeness. 

Recall that for all $p\in At, v(p)$ is an event, so $[p]$ is an event in $\Sigma$. Negation and intersection of events are events. By Lemmata~\ref{alles_events} and~\ref{Levent}, explicit knowledge, awareness, and implicit knowledge of events are also events, respectively. Thus, for every $\varphi \in \mathcal{L}_{\mathsf{At}}$, $[\varphi]$ is an event.

Proposition \ref{prop: semantic equivalence of K - L and A} shows that in complemented HMS models, $K_i(E) = L_i(E) \cap A_i(E)$, for any event $E \in {\Sigma}$, so the semantics of $\mathcal{L}_{\mathsf{At}}$ provided above immediately implies that:

\begin{prop}\label{prop: k iff L&A} For any complemented HMS model $\overline{\mathsf{M}}$, $\w \in \Omega$, $\varphi \in \mathcal{L}_{\mathsf{At}}$, and $\Psi \subseteq \mathsf{At}$ with $\mathsf{At}(\varphi) \subseteq \Psi$,
$\overline{\mathsf{M}},\w_{\Psi} \vDash k_i \varphi \leftrightarrow (\ell_i \varphi \wedge a_i \varphi).$
\end{prop}

An FH model and its HMS transform satisfy the same formulas in the language $\mathcal{L}_{\mathsf{At}}$ with implicit knowledge, explicit knowledge, and awareness as long as these formulas are defined at the corresponding states of the HMS transform. 

\begin{prop}\label{prop: equivalence2} For any FH model $\mathsf{K}_{\mathsf{At}}$ and its HMS transform $HMS(\mathsf{K}_{\mathsf{At}})$, 
for all $w \in W_{\mathsf{At}}$, $\varphi \in \mathcal{L}_{\mathsf{At}}$, and $\Phi \subseteq \mathsf{At}$ with $\mathsf{At}(\varphi) \subseteq \Phi$, $\mathsf{K}_{\mathsf{At}}, w \Vdash \varphi$ if and only if  $HMS(\mathsf{K}_{\mathsf{At}}), w_\Phi \vDash \varphi.$
\end{prop}

We now show that any complemented HMS model and its FH transform satisfy the same formulas from the language $\mathcal{L}_{\mathsf{At}}$ with implicit knowledge, explicit knowledge, and awareness.  

\begin{prop}\label{prop: equivalence1} For any complemented HMS model $\overline{\mathsf{M}}$ and its FH transform $FH(\overline{\mathsf{M}})$, for all $\varphi \in\mathcal{L}_{\mathsf{At}}$ and all $\w \in S_{At}$,
$\overline{\mathsf{M}}, \w \vDash \varphi$ if and only if $FH(\overline{\mathsf{M}}),\w \Vdash \varphi.$
\end{prop}

\section{Implicit Knowledge-based HMS Models and FH Models\label{sec: implicitHMS_transformations}} 

In this section, we focus on the relationship between implicit knowledge-based HMS and FH models. This relationship is even simpler than between HMS and FH models since implicit knowledge-based HMS models are arguably already closer to FH models. This is due to taking implicit knowledge and the awareness functions as primitives. 

\subsection{From FH Models to Implicit Knowledge-based HMS Models}\label{sec: u*-transf}

We can define a version of HMS transformation that is ``truncated'' after the $T$-transformation. It just keeps the first two steps of the HMS transformation. 

\begin{defin} For any FH model $\mathsf{K}_{\mathsf{At}}$, the \emph{truncated HMS transform} $HMS^*(\mathsf{K}_{\mathsf{At}}) = \langle I, \{S_{\Phi}\}_{\Phi \subseteq \mathsf{At}}, \\ (r^{\Phi}_{\Psi})_{\Psi \subseteq \Phi \subseteq \mathsf{At}}, (\Lambda^*_i)_{i \in I}, (\alpha^*_i)_{i \in I}, v \rangle$ is defined the first two steps of the HMS transform. 
\end{defin}

From Proposition~\ref{prop: T transform} follows now immediately: 

\begin{cor}\label{prop: trunctated HMS transform} For any FH model $\mathsf{K}_{\mathsf{At}}$, the truncated HMS-transform $HMS^*(\mathsf{K}_{\mathsf{At}})$ is an implicit knowledge-based HMS model.
\end{cor}

\subsection{From Implicit Knowledge-based HMS Models to FH Models}\label{sec: fh*-transf}

\begin{defin} For any implicit knowledge-based HMS model $\overline{\mathsf{M}}^* = \langle I, \{S_{\Phi}\}_{\Phi \subseteq \mathsf{At}}, (r^{\Phi}_{\Psi})_{\Psi \subseteq \Phi \subseteq \mathsf{At}}, (\Lambda_i)_{i \in I}, \\ (\alpha_i)_{i \in I}, v \rangle$, the  \emph{FH$^*$-transform} $FH^*(\overline{\mathsf{M}}^*) = \langle I, W_{\mathsf{At}}, (R_{\mathsf{At}, i})_{i \in I}, (\mathcal{A}_{\mathsf{At},i})_{i \in I}, V_{\mathsf{At}}\rangle$ is defined like the $FH$ transform except that the clause for the awareness correspondence is replaced by for any $i \in I$, 
\begin{itemize} 
\item $\mathcal{A}_{\mathsf{At}, i}: W_{\mathsf{At}} \longrightarrow 2^{\mathcal{L}_{\mathsf{At}}}$ is such that $\mathcal{A}_{\mathsf{At}, i} (\w) = \mathcal{L}_{\Phi}$ for $\Phi \subseteq \mathsf{At}$ with $\alpha_i(\w) = S_{\Phi}$. 
\end{itemize}
\end{defin}

The FH$^*$ transform indeed transforms any implicit knowledge-based HMS model into a FH model. 

\begin{prop}\label{prop: transform1*} For every implicit knowledge-based HMS model $\overline{\mathsf{M}}^*$, $FH^*(\overline{\mathsf{M}}^*)$ is a FH model for $\mathsf{At}$.
\end{prop}

\subsection{Equivalence of Implicit Knowledge-based HMS and FH Models} 

\begin{defin}\label{def: satisfiability of L2} Satisfaction of $\mathcal{L}_{\mathsf{At}}$ formulas in an implicit knowledge-based HMS model $\mathsf{M}^*$ is given like for complemented HMS models except that we have $\mathsf{M}^*, \w \vDash a_i \varphi$ if and only if $\alpha_i(\w) \succeq S([\varphi])$.
\end{defin} 

An FH model and its truncated HMS transform satisfy the same formulas in the language $\mathcal{L}_{\mathsf{At}}$ with implicit knowledge, explicit knowledge, and awareness with the provision that these formulas are defined at the corresponding states of the implicit knowledge-based HMS transform. This follows directly from the proof of Proposition~\ref{prop: equivalence2}.

\begin{cor}\label{prop: equivalence2*} For any FH model $\mathsf{K}_{\mathsf{At}}$ and its HMS transform $HMS(\mathsf{K}_{\mathsf{At}})$, 
for all $w \in W_{\mathsf{At}}$, $\varphi \in \mathcal{L}_{\mathsf{At}}$, and $\Phi \subseteq \mathsf{At}$ with $\mathsf{At}(\varphi) \subseteq \Phi$, $\mathsf{K}_{\mathsf{At}}, w \Vdash \varphi$ if and only if $HMS(\mathsf{K}_{\mathsf{At}}), w_\Phi \vDash \varphi.$
\end{cor}

Any implicit knowledge-based HMS model and its FH$^*$ transform satisfy the same formulas from the language $\mathcal{L}_{\mathsf{At}}$ with implicit knowledge, explicit knowledge, and awareness. 

\begin{prop}\label{prop: equivalence1*} For any implicit knowledge-based HMS model $\mathsf{M}^*$ and its FH$^*$ transform $FH^*(\mathsf{M}^*)$, for all $\varphi \in\mathcal{L}_{\mathsf{At}}$ and all $\w \in S_{At}$,
$\mathsf{M}^*, \w \vDash \varphi$ if and only if $FH^*(\mathsf{M}^*),\w \Vdash \varphi.$
\end{prop}

\section{Logic of Propositional Awareness\label{soundness_completeness}}

In the penultimate section, we explore the implications of the prior sections for axiomatizations of both the category of FH models and the complemented HMS models. In particular, we show that the Logic of Propositional Awareness is sound and complete with respect to the class of complemented HMS models. This is the first axiomatization of HMS models that feature also the notion of implicit knowledge. Previous axiomatizations of HMS models (Heifetz, Meier, and Schipper, 2008, Halpern and R\^{e}go, 2008) were confined to explicit knowledge and awareness only. We also show that the Logic of Propositional Awareness is sound and complete with respect to the class of implicit knowledge-based HMS models. This is the first axiomatization of implicit knowledge-based HMS models. Finally, it is also sound and complete with respect to the class of \emph{categories} of FH models. 
	
\begin{defin} The logic LPA is the smallest set of $\mathcal{L}_{\mathsf{At}}$ formulas that contains the axioms and is closed under the inference rules as follows:
All substitution instances of propositional logic, including 
$\top$ \setlength{\columnsep}{17pt}
\vspace{-10pt}
\begin{multicols}{2}
\noindent$(\ell_i\varphi\wedge(\ell_i\varphi\rightarrow \ell_i\psi))\rightarrow \ell_i\psi$\\
$k_i\varphi\leftrightarrow(\ell_i\varphi\wedge a_i\varphi)$\\
$a_i(\varphi\wedge\psi)\leftrightarrow(a_i\varphi\wedge a_i\psi)$\\
$a_i\neg\varphi\leftrightarrow a_i\varphi$\\
$a_ik_j\varphi\leftrightarrow a_i\varphi$\\
$a_ia_j\varphi\leftrightarrow a_i\varphi$\\
$a_i\ell_j\varphi\leftrightarrow a_i\varphi$\\
$a_i\varphi\rightarrow \ell_ia_i\varphi$\\
$\neg a_i\varphi\rightarrow \ell_i\neg a_i\varphi$\\
From $\varphi$ and $\varphi\rightarrow\psi$, infer $\psi$\\
From $\varphi$ infer $\ell_i\varphi$\\
$\ell_i\varphi\rightarrow\varphi$\\
$\ell_i\varphi\rightarrow \ell_i\ell_i\varphi$\\
$\neg \ell_i\varphi\rightarrow \ell_i\neg \ell_i\varphi$
\end{multicols}
\end{defin}
		
Recall that in a Kripke model or FH model, a formula is valid if it is true in every state. However, a formula $\varphi \in \mathcal{L}_{\mathsf{At}}$ may not even be defined at states of the FH model $\mathsf{K}_{\Psi}$ with $\mathsf{At}(\varphi) \nsubseteq \Psi$. Similarly, as we remarked earlier when introducing the semantics for HMS models, a formula may not be defined in every state of a HMS model. We say that $\varphi$ is defined in state $\omega$ in the complemented HMS model $\overline{\mathsf{M}}$ if $\omega \in \bigcap_{p \in \mathsf{At}(\varphi)} (v(p) \cup \neg v(p))$ (and analogously for implicit knowledge-based HMS models). Similarly, we say that $\varphi$ is defined in the FH model $\mathsf{K}_{\Psi}$ if $\mathsf{At}(\varphi) \subseteq \Psi$. 

Now we say that a formula $\varphi$ is valid in the complemented HMS model $\overline{M}$ if $\overline{M}, \omega \vDash \varphi$ for all $\omega$ in which $\varphi$ is defined  (and analogously for the implicit knowledge-based HMS model). Similarly, we say that $\varphi$ is valid in the category of FH models $\mathcal{C}(\mathsf{K}_{\Psi})$ if $\mathsf{K}_{\Psi}, w \Vdash \varphi$ for all $w \in W_{\Psi}$ for all $\mathsf{K}_{\Psi}$ in $\mathcal{C}(\mathsf{K}_{\mathsf{At}})$ for which $\varphi$ is defined. A formula is valid in a class of complemented HMS models $\mathcal{M}$ if it is valid in every complemented HMS model of the class  (and analogously for the class of implicit knowledge-based HMS models). A formula is valid in a class of categories of FH models $\frak{C}$ if it is valid in every category of FH models of the class. 

A proof in an axiom system consists of a sequence of formulae, where each formula in the sequence is either an axiom in the axiom system or follows from the prior formula in the sequence by an application of an inference rule of the axiom system. A proof of a formula $\varphi$ is a proof where the last formula of the sequence is $\varphi$. A formula $\varphi$ is provable in an axiom system, if there is a proof of $\varphi$ in the axiom system. An axiom system is sound for the language $\mathcal{L}_{\mathsf{At}}$ w.r.t. a class of complemented HMS models $\mathcal{M}$ if every formula in $\mathcal{L}_{\mathsf{At}}$ that is provable in the axiom system is valid in every complemented HMS model of the class $\mathcal{M}$ (and analogously for the class of implicit knowledge-based HMS models). Similarly, an axiom system is sound for the language $\mathcal{L}_{\mathsf{At}}$ w.r.t. a class of categories of FH models $\frak{C}$ if every formula in $\mathcal{L}_{\mathsf{At}}$ that is provable in the axiom system is valid in every category of FH models of the class $\frak{C}$. An axiom system is complete for the language $\mathcal{L}_{\mathsf{At}}$ w.r.t. a class of complemented HMS models $\mathcal{M}$ if every formula in $\mathcal{L}_{\mathsf{At}}$ that is valid in $\mathcal{M}$ is provable in the axiom system (and analogously for the class of implicit knowledge-based HMS models). Similarly, an axiom system is complete for the language $\mathcal{L}_{\mathsf{At}}$ w.r.t. a class of categories of FH models $\frak{C}$ if every formula in $\mathcal{L}_{\mathsf{At}}$ that is valid in $\frak{C}$ is provable in the axiom system.

\begin{cor}\label{class of categories axiomatization} LPA is sound and complete w.r.t.
\begin{enumerate} 
\item the class of categories of FH models,
\item the class of complemented HMS models, 
\item the class of implicit knowledge-based HMS models.
\end{enumerate}   
\end{cor}

Fagin and Halpern (1988), Halpern (2001), and Halpern and R\^{e}go (2008) claim that LPA is sound and complete w.r.t. the class of FH models. The proof of 1. now follows from invariance of modal satisfaction relative to sublanguages between FH models in each category of FH models, i.e., Proposition~\ref{epistemic_equivalence}. The proof of 2. follows from Propositions~\ref{prop: equivalence2} and~\ref{prop: equivalence1}. The proof of 3. follows from Corollary~\ref{prop: equivalence2*} and Proposition~\ref{prop: equivalence1*}.

\section{Discussion}

The constructions also allowed us to consider the relation between FH models and HMS models, not just with respect to explicit knowledge and awareness as in the prior literature but also with respect to implicit knowledge. We show modal equivalence between FH and HMS models by transforming one model into another. Each model and its transform satisfy the same formulae from a language of implicit, explicit knowledge and awareness. This equivalence is used to show that the Logic of Propositional Awareness is sound and complete for the class of HMS models. Compared to the prior literature, this axiomatization is now for a language that also features implicit knowledge. 
\begin{figure}\caption{Relations between Approaches to Awareness\label{relations}}
\begin{center}
\includegraphics[scale = 0.15]{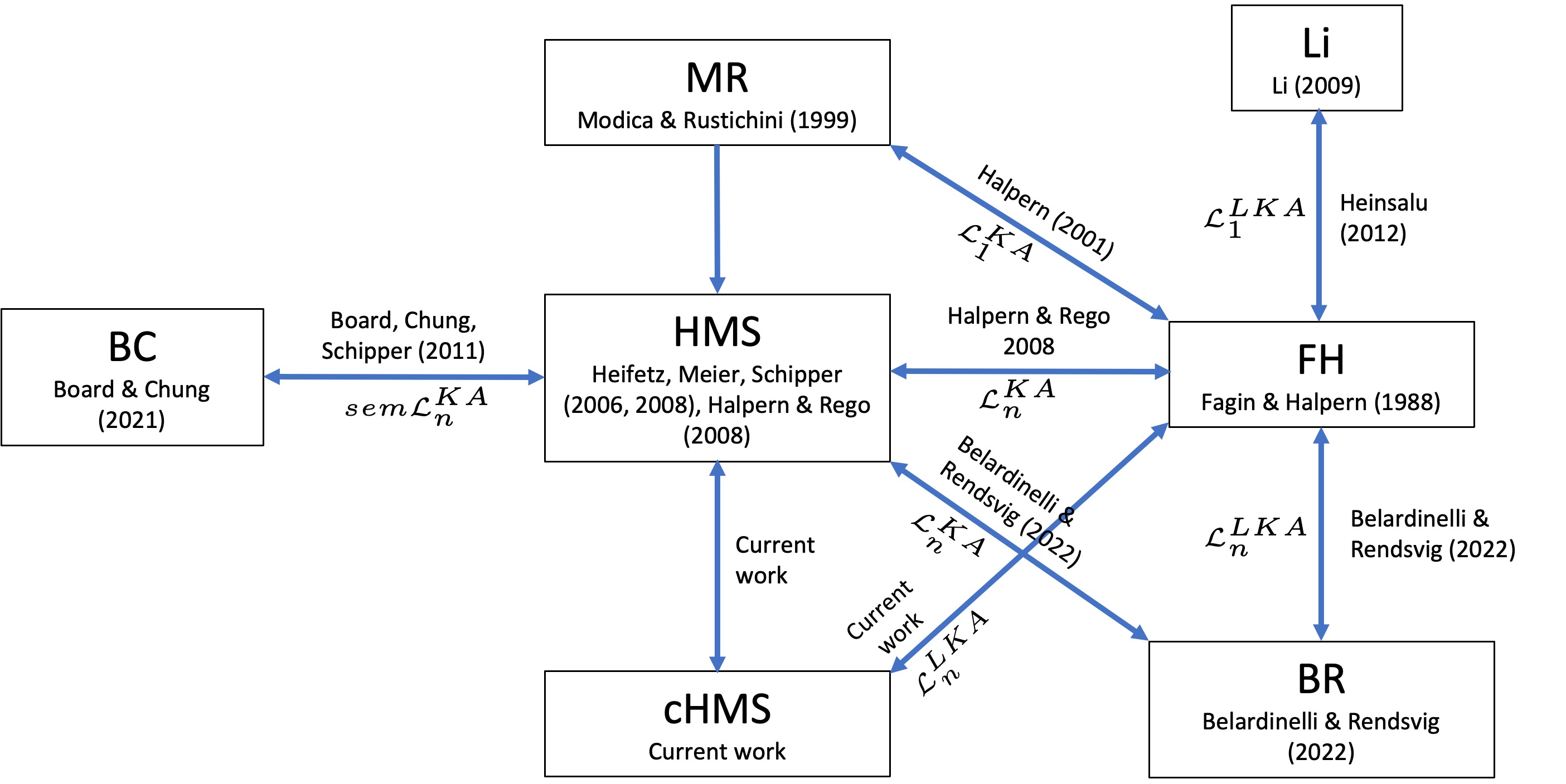} 
\end{center}
\end{figure}

The relations between various models of awareness in the literature are depicted in Figure~\ref{relations}. Beside FH models of Fagin and Halpern (1988) and HMS models of Heifetz, Meier, and Schipper (2006, 2008), we consider generalized standard models by Modica and Rustichini (1999), information structures with unawareness by Li (2009), object-based unawareness models by Board and Chung (2021), and Kripke lattices by Belardinelli and Rendsvig (2022). Equivalences hold for various languages also shown in the figure. We indicate the implicit, explicit, and awareness modality by superscripts $L$, $K$, and $A$, respectively. Some structures like Modica and Rustichini (1999) and Li (2009) feature just a single agent. We indicate this with the subscript ``$1$'' for single agent and ``$n$'' for multiple agents. For instance, $\mathcal{L}_n^{L, K, A}$ is the language featuring multiple agents, implicit knowledge, explicit knowledge, and awareness. The equivalence between Board and Chung (2021) is shown only at the level of semantics, i.e., at the level of events. The relation between Modica and Rustichini (1999) and Heifetz, Meier, and Schipper (2008) indicates that latter axiomatization can be seen as a multi-agent version of former. All shown relations pertain to rich structures featuring partitional knowledge and awareness generated by primitive propositions.

Recently, Schipper (2022) extended HMS models to awareness of unawareness by introducing quantified events. It would be straightforward to complement his structure with implicit knowledge as defined in the current work. Agents could then reason about the existence of their implicit knowledge that they are not aware of. Such reasoning bears some similarity to the notion of speculative knowledge in van Ditmarsch et al. (2018). Awareness-of-unawareness structures with implicit knowledge would also allow for a better comparison to awareness structures with quantification of formulae for modeling reasoning about knowledge of unawareness (Halpern and R\^ego, 2009, 2012), object-based unawareness (Board and Chung, 2021), and quantified neighborhood structures with awareness (Sillari, 2008).

%Bibliography.....................................................
	
\bibliographystyle{eptcs}

\providecommand{\urlalt}[2]{\href{#1}{#2}}
\providecommand{\doi}[1]{doi:\urlalt{http://dx.doi.org/#1}{#1}}

\end{document}